# Absence of lateral domain wall mobility in $Zn_{1-x}Mg_xO$ thin films

Jack Eckstein,[1] Kyle P. Kelley,[1] William Prudnick,[2] Jon-Paul Maria,[2] Stephen Jesse,[1]

Neus Domingo,[1] Rama K. Vasudevan[1]*

[1] Center for Nanophase Materials Sciences, Oak Ridge National Laboratory, Oak Ridge, 37831, Tennessee, USA.

[2] Department of Material Science and Engineering, The Pennsylvania State University, University Park, PA 16802, USA.

## ABSTRACT

Polarization reversal in ferroelectrics arises from the coupled processes of domain nucleation and subsequent growth, yet the governing mechanisms differ fundamentally between classical perovskite oxides and emerging wurtzite ferroelectrics. While switching in perovskites is typically governed by mobile domain walls whose field-driven propagation dominates macroscopic kinetics, here we show that polarization reversal in wurtzite $Zn_{1-x}Mg_xO$ proceeds through a qualitatively different pathway. Using scanning oscillator microscopy, in combination with point pulse-imaging methods, we directly map local switching events and domain wall responses, revealing that domain walls in $Zn_{1-x}Mg_xO$ exhibit negligible lateral mobility (<10nm) and that polarization reversal proceeds predominantly through the nucleation of vertically extended columnar filaments with a lateral size on the order of the grains. This nucleation-controlled switching contrasts sharply with the growth-mediated dynamics characteristic of perovskite ferroelectrics and explains the abrupt, spatially localized switching behavior observed in wurtzite systems. These results establish nucleation-dominated filamentary reversal as a defining switching mechanism in $Zn_{1-x}Mg_xO$ and point towards the need for further studies to understand correlation lengths and nucleation processes across a range of grain sizes.

*vasudevanrk@ornl.gov

## Introduction

Oxide perovskite ferroelectrics established the modern framework for electric-field-driven domain nucleation and growth. Early measurements on $BaTiO_3$ demonstrated the characteristic field dependence of switching kinetics, forming the canonical picture of activated polarization reversal.[1] Subsequent electrostatic analysis revealed the so-called Landauer paradox: homogeneous nucleation of a reversed domain in a uniformly polarized crystal would require extremely large energies.[2]This insight motivated the now-standard view that switching proceeds through energetically favorable nucleus geometries and, in practice, through heterogeneous nucleation at defects, interfaces, or pre-existing structural inhomogeneities. At the microscopic scale, domain wall motion in perovskite ferroelectrics can occur through repeated nucleation of kinks or ledges that spread laterally, a "sidewise" growth mechanism originally formulated for 180◦ walls in BaTiO3.[3] At larger spatial and temporal scales, switching kinetics are often described using the Kolmogorov–Avrami framework adapted to ferroelectrics, which relates the switched volume fraction to idealized nucleation and growth rates.[4] Real thin films, however, frequently depart from this idealized behavior. Such deviations are commonly interpreted in terms of heterogeneous switching with spatially distributed local kinetics,[5] or in nucleation-limited scenarios where broad distributions of activation barriers and local fields render nucleation—rather than wall propagation—the rate limiting step.[6] Quantitative understanding of domain-wall energetics and mobility in perovskite ferroelectrics has emerged from combined atomistic and mesoscopic theory, bulk measurements, and direct imaging. First-principles calculations established wall structures, widths, and energies in prototypical systems such as $PbTiO_3$, providing a microscopic basis for models of mobility and pinning.[7] Experimentally, epitaxial PZT films exhibit thermally activated domain-wall creep, demonstrating that disorder controlled wall motion can govern domain expansion in realistic energy landscapes.[8] Advances in scanning-probe and in situ electron microscopy have further enabled real-time visualization of domain nucleation and propagation, for example in $BiFeO_3$ , directly imaging nucleation sites and wall morphology.[9] Together with quantitative decoupling of intrinsic and extrinsic responses in thin films[10] and studies of substrate-clamping effects on irreversible wall dynamics,[11] these results establish mobile domain walls as central actors in perovskite ferroelectric switching and the overall piezoresponse.

**Background**

Wurtzite ferroelectrics, most prominently Sc-alloyed AlN and $Zn_{1-x}Mg_xO$, place polarization reversal in a distinct regime where switching occurs within a tetrahedrally bonded, wide-bandgap lattice possessing only two polarization directions and large coercive fields compared to their perovskite cousins. The demonstration of robust ferroelectricity in $Al_{1-x}Sc_xN$ established a model material system in which nucleation and domain evolution can be systematically investigated in thin-film device geometries.[12] Subsequent studies across AlN-based systems, including AlScN and AlBN, revealed a pronounced temperature dependence of the coercive field while maintaining strong remanent polarization, indicating that switching is governed primarily by activation barriers rather than polarization magnitude.[13] Together, these observations suggest that the kinetics and energetics of polarity reversal in wurtzite ferroelectrics are dominated by barrier controlled nucleation processes that differ fundamentally from those traditionally emphasized in perovskite oxides. A major advance in understanding wurtzite switching has come from atomic-scale mechanistic observations. In-situ atomic resolution studies of $Al_{0.94}B_{0.06}N$ directly visualized polarity reversal proceeding through local structural rearrangements toward a transient nonpolar configuration, with first-principles calculations supporting an antipolar intermediate along the switching pathway.[14] These results imply that nucleation barriers in wurtzite ferroelectrics cannot be described solely by electrostatic and elastic considerations—historically central in perovskite switching models,[2,3] but must also incorporate local bonding topology and intermediate-state energetics. Such structurally mediated pathways provide a natural explanation for the frequently abrupt and cooperative nature of polarity reversal observed in nitride-based ferroelectrics. Consistent with these microscopic insights, switching transients in wurtzite ferroelectrics often exhibit unusually abrupt behavior at technologically relevant fields, motivating extensions of classical nucleation–growth theory. Models incorporating simultaneous non-linear nucleation and growth account for the large apparent Avrami exponents reported in AlScN systems by allowing substantial growth and impingement to occur while the nucleation rate is still increasing towards its maximum.[15] Device-level measurements of hysteresis and transient kinetics further highlight the importance of linking electrical signatures to real-space domain evolution while operating near breakdown-limited fields.[16] Beyond single-layer films, heterostructures introduce qualitatively new switching pathways: proximity ferroelectricity demonstrates that a nominally unswitchable polar layer can reverse when coupled to a switchable neighbor, consistent with interface-mediated antipolar nuclei and domain-wall-assisted barrier reduction.[17] Complementary thermodynamic treatments provide a Landau–Ginzburg framework describing proximity switching regimes and the roles of internal fields and charged defects. [18] A complementary frontier concerns the structure, energetics, and functionality of wurtzite domain walls themselves. Electric-field-induced domain walls in these systems can be charged and structurally reconstructed, with compensation mechanisms that produce reconfigurable electronic states and conductivity.[19] Spatially resolved *in-situ* measurements in epitaxial wurtzite films reveal strongly anisotropic switching behavior, with nucleation often localized near electrodes, followed by rapid lateral propagation along interfaces and comparatively slow vertical advance through the film thickness, frequently leaving persistent non-switching interfacial layers.[20] Such observations elevate interfaces, defect chemistry,[21] and electrostatic boundary conditions to first-order control parameters, including the formation of interfacial dead layers linked to defect–strain coupling.[22] Parallel advances in multiscale modeling, including data-driven interatomic potentials, now enable multidomain simulations in AlN-class systems beyond direct first-principles length scales.[23] Extending beyond nitride ferroelectrics, recent studies of $Zn_{1-x}Mg_xO$ demonstrate polarity-dependent switching kinetics and cycling-driven evolution of nucleation behavior, highlighting oxide wurtzites as an emerging platform for probing switching physics in this structural family.[24]

**Comparative analysis**:

Despite sharing the general phenomenology of nucleation followed by domain-wall-mediated growth, perovskite and wurtzite ferroelectrics differ in several mechanistically decisive respects. First, the switching unit and accessible variants are fundamentally different. Perovskites commonly host multiple ferroelastic and ferroelectric variants (e.g., 90◦ /180◦ in tetragonal systems and 71◦ /109◦ /180◦ in rhombohedral BiFeO3), enabling several low-energy reorientation pathways and complex domain topologies. In contrast, wurtzite ferroelectrics typically switch only between two polarity states along the $c$ axis,[12] constraining reversal to a more geometrically restricted pathway. Second, the origin of nucleation barriers reflects different dominant energy contributions. Classical perovskite treatments emphasize depolarization fields, electrostatic screening, and nucleus geometry,[2,3] while modern atomistic perspectives recognize that realistic nanoscale nuclei and defect-assisted pathways mitigate otherwise prohibitive barriers.

Multiscale simulations have further shown that traditional classical nucleus pictures for domain-wall motion can be energetically inconsistent, motivating corrected nucleation-and-growth mechanisms grounded in atomistic energetics.[25] In wurtzites, by contrast, atomic resolution observations demonstrate that nucleation barriers include a strong local structural component governed by bonding topology and intermediate-state energetics,[14] often coupled to interfacial constraints and dead-layer effects.[20,22] Together, these differences indicate that while perovskite switching barriers are often dominated by electrostatic and mesoscale considerations, wurtzite switching intrinsically couples electrostatics to local structural transformation pathways.

Third, growth kinetics and rate-limiting processes differ systematically between the two classes. In perovskites, domain-wall motion frequently exhibits disorder-controlled regimes such as creep,[8] and deviations from ideal Kolmogorov–Avrami–Ishibashi kinetics are commonly interpreted in terms of heterogeneous nucleation rates and distributed barrier landscapes.[5,6] Wurtzite ferroelectrics, conversely, often display unusually abrupt switching transients that challenge constant-nucleation assumptions and motivate models incorporating time-dependent nucleation concurrent with growth and impingement.[15,20] Spatially resolved measurements in epitaxial wurtzites further reveal strongly anisotropic switching behavior, with nucleation frequently localized near electrodes and propagation constrained by interfaces or geometric bottlenecks.[16] These observations suggest that switching in wurtzites may be governed less by bulk-like domain wall mobility and more by interface-controlled nucleation and propagation pathways. Defects and boundary conditions influence both material classes, but with different emphases. In perovskite thin films, microstructural length scales, elastic boundary conditions, and defect pinning largely determine the contribution of domain wall motion to macroscopic response and switching reversibility.[10,11] In wurtzites, defects and interfaces can control not only wall pinning but whether switching occurs at all prior to electrical breakdown.[21]

In this work, we leverage automated microscopy as well as the recently developed Scanning Oscillator microscopy (SO-PFM) technique[26] to probe complementary aspects of domain-wall stability and mobility in $Zn_{1-x}Mg_xO$. AEcroscopy[27] is used to implement an automated pulse–probe workflow in which a selected domain wall is electrically pulsed and subsequently re-measured over extended time intervals to quantify any post-pulse relaxation, drift, or recovery under controlled conditions. In parallel, SO-PFM is used to test real-time domain-wall mobility by applying a slowly varying, large-amplitude bias during scanning while continuously sensing the local PFM response with a superimposed high-frequency probe signal. This approach allows any lateral wall displacement to be detected directly as a function of instantaneous applied field. Together, time-resolved pulse–relaxation measurements and in situ oscillator-driven tracking provide complementary constraints on whether switching proceeds through lateral domain-wall motion or through repeated nucleation events.

**Results and Discussion**

We probe domain wall mobility in $Zn_{1-x}Mg_xO$ (ZnMgO) thin films using two complementary approaches, illustrated in Figure 1. Details of sample preparation are provided in the methods section. In the first case, shown in Figure 1d, automated pulse–probe experiments implemented on the AEcroscopy platform are used to apply voltage pulses of prescribed amplitude and duration to pre-existing domain walls through the AFM tip, enabling the influence of these walls on the nucleation and growth of reversed domains to be assessed directly. In the second, the recently developed scanning oscillator PFM (SO-PFM) technique (Figure 1b,c), a low-frequency (200–700 Hz), high-amplitude drive bias is superimposed on a low-amplitude (1–2 V), high-frequency (~10 kHz) detection signal applied to the tip. The responses at the two frequencies are monitored simultaneously, either with analog lock-in amplifiers or digitally through full-waveform acquisition (the general-mode, or G-mode[28], approach). Because the wall is imaged continuously while the drive bias sweeps through its extrema, any field-driven lateral wall displacement is detected directly as a function of the instantaneous applied field.

Polarization reversal in $Zn_{1-x}Mg_xO$ is believed to proceed through the independent-column, or "fringing ridge," mechanism illustrated in Figure 1a, in which a high density of nuclei forms at the surface or electrode interface and reversal propagates as independent columnar domains extending to the bottom electrode, rather than through lateral domain wall growth. This picture is supported by theoretical fits to macroscopic polarization switching transients, as well as by previous piezoresponse force microscopy and spectroscopy studies, especially for the nitride system.[13,24] Nevertheless, the behavior of the domain walls themselves, and the extent to which their lateral mobility contributes to switching remains poorly understood. Here, we directly interrogate the impact of pre-existing domain walls, and the mobility of those walls, on the switching process.

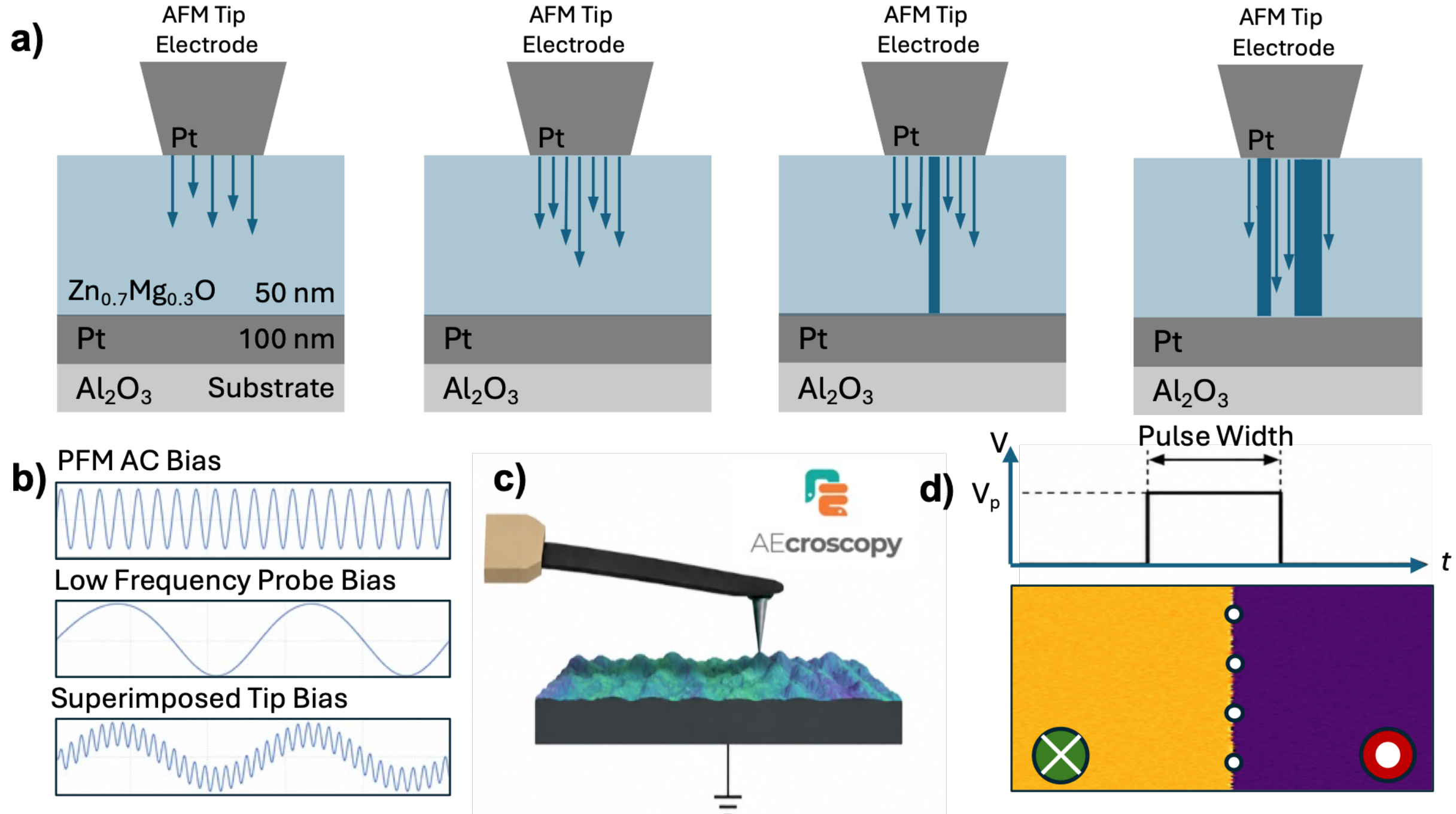

**Figure 1.** Experimental approaches for probing domain wall mobility in $Zn_{1-x}Mg_xO$. a) Fringing ridge (independent column) model of switching, in which a high density of nuclei at the electrode propagates as independent columns through the film thickness, without lateral wall growth. b) Scanning oscillator PFM (SO-PFM): a low-frequency, high-amplitude drive is mixed with a high-frequency, low-amplitude read signal, and the response at both frequencies is monitored simultaneously and applied to the tip as it is raster across the sample (c). (d) Automated pulse–probe experiments implemented on the AEcroscopy platform, in which pre-existing domain walls are pulsed by the AFM tip with prescribed voltages and pulse widths.

A ZMO film with x=0.30 was sputtered on Pt-coated Sapphire substrate; the film is textured along the c-axis, with a thickness of ~50nm and shows small grains (~5-10nm size) with more details available elsewhere.[29] SO-PFM measurements were performed on the ZMO film at four drive amplitudes, beginning in the sub-coercive regime (10 V) and increasing to amplitudes well above the coercive field, to establish whether the domain walls are mobile within this excitation range. Figure 2 shows the topography (top row), PFM amplitude (middle row), and PFM phase (bottom row) acquired after each measurement. At a 10 V drive, the pre-existing domain wall, written by application of a DC bias to the PFM tip, remains clearly visible, while no signature of the SO experiment that was performed is observed; that is, apparently no irreversible domain wall displacement was induced (to within resolution). We note that each SO-PFM image comprises several hundred consecutive line scans crossing the same wall, such that independent domain wall segments are sampled repeatedly over many drive cycles. Furthermore, frame-by-frame tracking of the instantaneous wall position in the in-situ SO response at 10 V drive (Figure S1), yields a bias-synchronous apparent displacement of ~8 nm, below the 10 nm pixel size of the measurement.

At 20 V, localized switching is apparent in several regions, evidenced by reduced amplitude in the lower half of the switched areas and by fluctuations in the phase signal; the pre-existing wall, however, remains unperturbed. This behavior persists to the highest drive amplitudes: no bias within the accessible range is sufficient to irreversibly displace the wall. Our measurement places an upper bound of approximately 8 nm on any reversible, field-driven wall displacement in the sub-coercive regime (S1).

This result is striking when contrasted with $PbTiO_3$, for which SO-PFM measurements resolved reversible, hysteretic domain wall displacement above an oscillator amplitude threshold of only ≈65kV/cm, with creep-like wall dynamics at drive amplitudes of 75-200kV/cm.[26] In $Zn_{1-x}Mg_xO$, by contrast, no wall response is detected at drive amplitudes nearly an order of magnitude larger (~4MV/cm). Moreover, once the applied field exceeds the coercive field, irreversible modification of the domain structure would be expected and would be evident in the PFM images acquired after the measurement. The absence of any such changes, even at the highest drive amplitude, indicates that the energy barrier for lateral motion of written domain walls in $Zn_{1-x}Mg_xO$ is extremely high. Polarization reversal instead proceeds predominantly through nucleation and growth along the vertical (polar) direction, with lateral wall propagation strongly suppressed and unresolvable within the spatial resolution of the microscope.

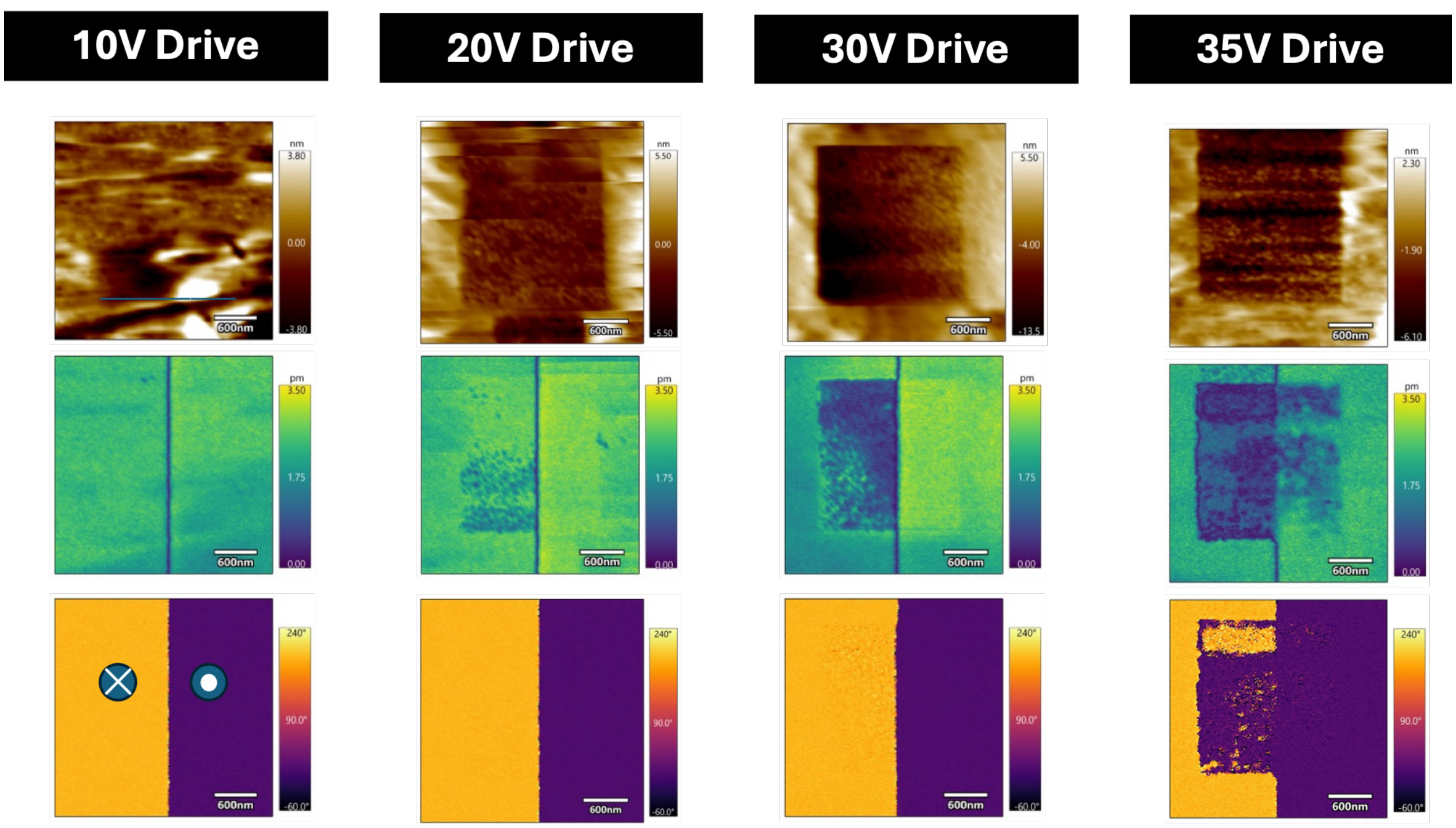

**Figure 2.** SO-PFM measurements on $Zn_{0.7}Mg_{0.3}O$ thin films at drive amplitudes of 10, 20, 30, and 35 V (left to right). Topography (top row), PFM amplitude (middle row), and PFM phase (bottom row) acquired after each SO-PFM

measurement. The pre-existing domain wall, written by a DC bias applied to the tip, remains unperturbed at all drive amplitudes, indicating the absence of irreversible lateral domain wall motion.

Although the SO-PFM measurements show no evidence of lateral wall motion, pre-existing walls could nonetheless influence the nucleation and growth of new domains. To test this possibility, a domain wall was first written by switching a selected region to the up-polarized state, and voltage pulses of varying amplitude were then applied at positions along the newly formed wall. Figure 3 shows the vertical PFM phase and amplitude after pulses of 40, 50 and 60V (equivalent for 50nm film, 8MV/cm, 10MV/cm and 12MV/cm in an electrode geometry) applied adjacent to the wall, together with magnified views of the boxed region in the amplitude image. For pulse amplitudes above approximately 50V, a new domain forms as an apparent expansion of the pre-existing domain, observed in both the amplitude and phase channels. Even at the highest pulse amplitude, however, unswitched regions persist at the periphery of the reversed area, as highlighted in the insets. This morphology is inconsistent with field-driven motion of the pre-existing wall; rather, the tip field nucleates new domains and drives polarization reversal in the conventional manner for ZMO, i.e., apparently independent of the pre-existing wall.

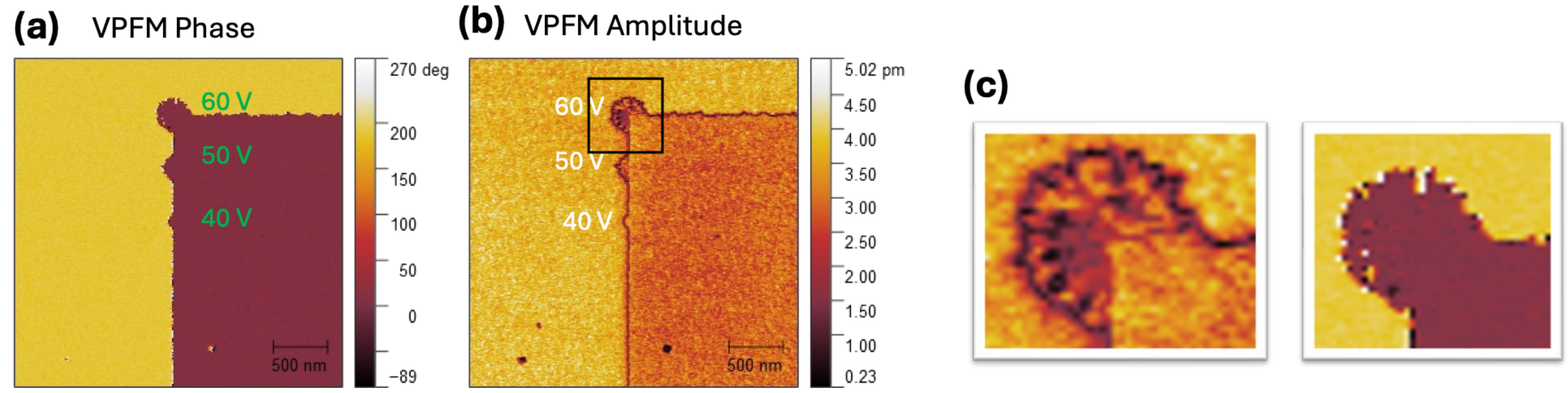

**Figure 3.** Effect of a pre-existing domain wall on switching. Vertical PFM phase (a) and amplitude (b) after pulses of 40, 50, and 60 V applied at positions along a written domain wall. (c) Magnified amplitude (left) and phase (right) views of the boxed region in the amplitude image in (b), showing unswitched regions persisting at the periphery of the newly switched domain.

To gain qualitative insight into the switching kinetics, we constructed a minimal model of ferroelectric switching under the following assumptions: we neglect elastic and depolarization effects and long-range electrostatics. The film comprises grains of mean size 10 nm; and switching proceeds via nucleation followed by instantaneous vertical growth with zero lateral growth (the “independent column switching” model). The electric field produced by the tip is modeled as a Lorentzian:

$$E(r) = \frac{E_0}{1 + \left(\frac{r}{r_0}\right)^2}$$

where $r_0$ = 50 nm and $E_0$ is expressed in units of the coercive field. The film is treated as polycrystalline, with grain centers generated by a Poisson process followed by Voronoi tessellation with a mean grain size of 10 nm, and each grain is assigned a nucleation barrier drawn from a lognormal distribution:

$$E_{n,g} = E_{n,med} \exp(\sigma \xi_g)\,, \xi_g \sim \mathcal{N}(0,1)$$

where $E_{n,med}$ is the median nucleation threshold field and $\sigma$ controls the disorder strength. We set the nucleation threshold constant within each grain. Switching is treated as kinetic in nature, with nucleation at a site resulting in complete, instantaneous reversal of that pixel in the simulation. Although idealized, this assumption is motivated by the fast columnar switching observed experimentally and by the comparatively slow timescale of PFM detection. The nucleation rate within each grain follows a Merz-type activation law:

$$I(\boldsymbol{r}) = I_0 \exp\left[-\left(\frac{E_n(\boldsymbol{r})}{|E(\boldsymbol{r})|}\right)\right]$$

where $I_0 = 5\times10^6$ in our simulations. For comparison with a conventional ferroelectric, wall motion is additionally modeled as:

$$v(r) = v_0 \exp\left[-\left(\frac{E_v}{|E(r)|}\right)\right]$$

where $E_v$ is the activation field for wall motion. For a constant-field pulse of duration $T$, the switching probability at location $r$ is:

$$p(\boldsymbol{r}) = 1 - \exp[-I(\boldsymbol{r})T]$$

The simulation results are shown in Figure 4, in which lateral wall motion can be selectively enabled or disabled. With wall motion disabled, the simulations reproduce the key experimental observations of Figure 3. That is, we observe multiple grains that remain unswitched within the reversed region. However, when we enable domain wall motion, nearly all grains switch and a compact, near-circular domain emerges, as expected for a conventional ferroelectric. These results further support the conclusion that lateral domain wall motion is insignificant in these $Zn_{1-x}Mg_xO$ thin films, with switching instead governed by grain-resolved nucleation.

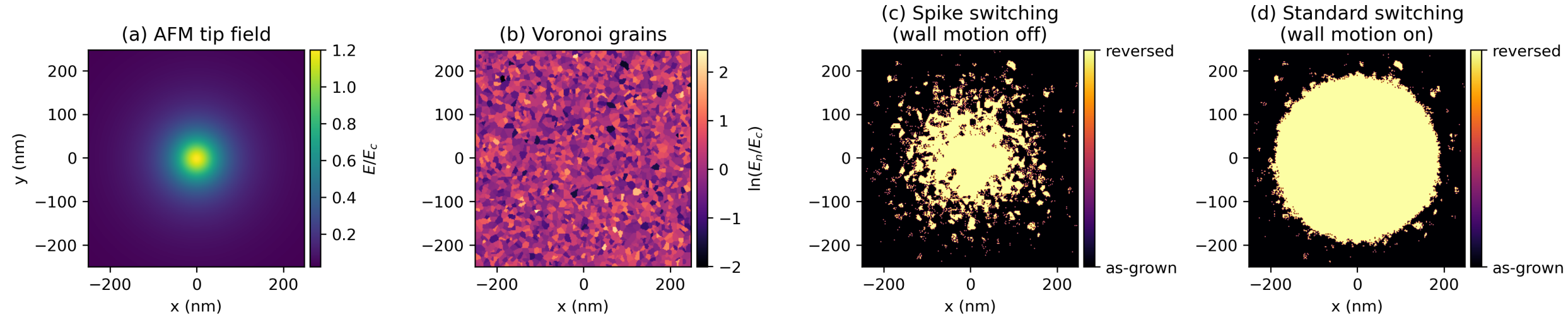

**Figure 4.** Spike switching model simulations. (a) AFM tip field $E/E_c$; (b) map of the grain-resolved nucleation barriers $\log(E_n/E_c)$ generated by Voronoi tessellation; (c) final domain state for spike switching (lateral wall motion off), reproducing the unswitched grains observed experimentally; and (d) standard switching (wall motion on), yielding a compact, near-circular switched domain.

## Discussion

A central open question is whether switching in wurtzite ferroelectrics can be organized within a unified nucleation-limited versus growth-limited taxonomy analogous to that developed for perovskite thin films.[5,6] Thickness-resolved imaging in epitaxial films indicates electrode-proximal nucleation and strongly anisotropic propagation,[20] yet a predictive framework that quantitatively links atomistic switching intermediates,[14] interfacial dead-layer energetics,[22] and macroscopic transient signatures[15,16] remains lacking. Establishing such connections would enable switching kinetics in wurtzites to be interpreted within a physically grounded hierarchy spanning atomic-scale pathways to device-scale observables.

A second open challenge is resolving the nucleation and "independent-column" character of the switching process. Atomistic studies of wurtzite ferroelectrics indicate that polarization reversal proceeds sequentially through a transient antipolar state, facilitated by the large local strain fluctuations that Mg substitution introduces into the tetrahedral bonding network.[14] PFM measurements lack the resolution to image this directly, and the correlation length of the switched regions is on the order of the grain size in these sputtered films (Figs. 2, 3). This is suggestive of nucleation at grain boundaries or other local heterogeneities; for example, charged defects or injected carriers segregating to grain boundaries could locally lower the nucleation barrier and seed reversal. We emphasize that this grain-boundary-nucleation scenario is speculative and warrants dedicated theoretical and experimental study.

## Supplementary Material

Supplementary material is provided with this manuscript that includes additional analysis of the scanning oscillator data.

**Acknowledgments**

The PFM experiments, sample generation and modeling were demonstrated and validated with support by the Center for 3D Ferroelectric Microelectronics Manufacturing (3DFeM2), an Energy Frontier Research Center funded by the US Department of Energy, Office of Science, Office of Basic Energy Sciences Energy Frontier Research Centers program under award DE-SC0021118. Scanning oscillator development was supported by the Center for Nanophase Materials Sciences (CNMS), which is a US Department of Energy, Office of Science User Facility at Oak Ridge National Laboratory.

**Conflicts of Interest**

The authors declare no conflicts of interest.

**Methods**

A $Zn_{0.7}Mg_{0.3}O$ thin film was prepared via sputter-down reactive RF magnetron co-sputtering on platinized c-plane (0001) sapphire substrates. 2"-diameter metallic zinc (Kurt J. Lesker; 99.995%) and metallic magnesium (Kurt J. Lesker; 99.95%) targets were affixed to cathodes ~23° incident to the substrate approximately 6 cm to the stage. Ozone-oxygen (O3/O2) gas (~10% w/v ozone; ~90% diatomic oxygen) was synthesized from ultra-high purity (UHP) oxygen gas using an Oxidation Technologies, LLC. ATL-30 ozone generator. 18 sccm argon was flowed as a sputtering gas and 6 sccm of O3/O2 was flowed as a reactant gas and total pressures during deposition were 0.75mTorr. PFM measurements were performed at room temperature on an Oxford Instruments Cypher Vero AFM via the interferometric detection system and positioning of the laser in such a way as to minimize the first resonance mode when in contact (to ensure minimization of long-range electrostatic forces).

## SUPPORTING INFORMATION

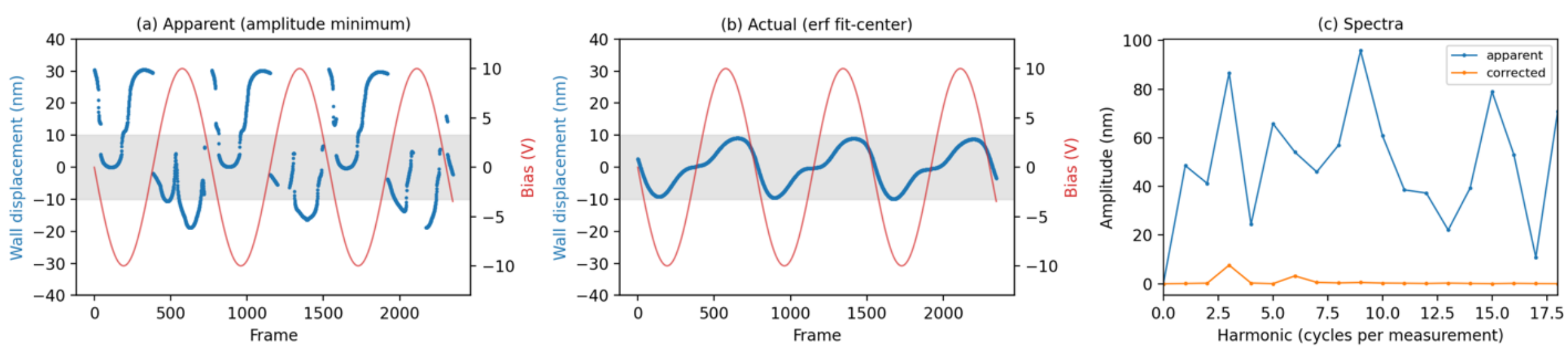

**Figure S1.** In-situ domain-wall tracking during SO-PFM at 10 V (a) Apparent wall position from the response-amplitude minimum (blue) versus oscillator bias (red); the shaded band denotes ±1 pixel (10 nm). (b) Electrostatically corrected wall position. The corrected position varies in phase with the bias with an amplitude of only ~8 nm, remaining within approximately one pixel throughout. (c) Displacement spectra of the wall for the two methods, showing that the apparent domain wall position contains multiple harmonics, whereas the corrected wall position contains very little signal.

In Figure S1 we show the results (a-c) from the Scanning Oscillator measurement before and after the fitting procedure to subtract the electrostatic contributions (as detailed in Raghuraman et al. [26]. Specifically, in this experiment, the SO results were obtained using an oscillation frequency of ~650Hz, and an AC amplitude of 10V. The complex piezoresponse was averaged over ~200 lines of the domain wall and we chose 3 representative periods of the oscillation. In (a), we show the conventional "apparent" domain position, taken as the minimum of the response amplitude. In (b), we plot the more robust position obtained by rotating the complex response to maximize the piezoelectric channel and fitting an error function whose center defines the wall location.

The two methods behave differently, in both sign and magnitude, as was also the case in [26]. The apparent wall position moves opposite to that expected for genuine field-driven wall displacement and is a signature of electrostatic artifacts, rather than physical wall motion [26]. In contrast, the corrected domain wall position varies in phase with the bias but with an amplitude of only ~8 nm, which is below the 10 nm pixel size. The displacement spectrum of the apparent position carries substantial power at the drive frequency and its harmonics, whereas the corrected position is featureless at the same scale.

This suggests the apparent bias-synchronous wall motion in this ZMO film is dominated by electrostatic contrast modulation, and that the true wall position is stationary to within the resolution of the measurement (~10 nm).